\documentclass[aps, prd, twocolumn, nofootinbib]{revtex4-1}

\usepackage[top=1in, bottom=1in, left=1in, right=1in]{geometry}

\usepackage{amsmath}
\usepackage{amsfonts}
\usepackage{wasysym}
\usepackage{graphicx}
\usepackage{comment}
\usepackage{tensor}

\def\filetype{pdf}
\def\path{}

\begin{document}

%---TITLE PAGE------------------------------------------------------------------
\title{Type II critical behavior of gravitating magnetic monopoles}
\author{Ben Kain}
\affiliation{Department of Physics, College of the Holy Cross, Worcester, Massachusetts 01610, USA}

\begin{abstract}
\noindent I study type II critical collapse in the spherically symmetric gravitating magnetic monopole system.  This is an Einstein-Yang-Mills-Higgs system with two matter fields:\ a field parametrizing the scalar field gauged under $SU(2)$ and a field parametrizing the gauge field.  This system offers interesting differences compared to what is commonly found for type II collapse in other systems.  For example, instead of the critical solution sitting between collapse and complete dispersal of the matter fields, on the non-black hole side of the critical solution, the matter fields settle down to a static and stable configuration.  More interesting, however, is that I find strong evidence for the existence of two critical solutions, each with their own set of scaling and echoing exponents, which I determine numerically.
\end{abstract}

\maketitle
%\tableofcontents

%============================================

\section{Introduction}

Critical gravitational phenomena was first discovered by Choptuik \cite{Choptuik:1992jv} in the form known as type II.  In type II, a spacetime evolves such that it eventually leads to the formation of a black hole or does not.  For example, consider regular initial data, parametrized by a single parameter, $p$, such that for $p>p^*$ the spacetime dynamically evolves from the initial data to a spacetime containing a black hole, while for $p<p^*$ a black hole does not form and the matter fields disperse to infinity.  The spacetime with $p=p^*$ is called the critical solution and the remarkable behavior at or near $p^*$ is what is meant by type II critical phenomena.

What is remarkable is that near-critical spacetimes, i.e.\ spacetimes for which $p$ is near $p^*$, exhibit self-similarity.  In the case of discrete self-similarity, a scale invariant function, $Z$, obeys
\begin{equation}
Z( \tau+\Delta, \ln r + \Delta) = Z(\tau, \ln r),
\end{equation}
where the echoing exponent, $\Delta$, is universal, in that it is independent of initial data.  In the above equation $\tau = \ln(T^* - T)$, where $T$ is the central proper time (i.e.\ the proper time at the origin) and $T^*$ is a constant called the accumulation time.  Also remarkable is that the mass of the black hole at collapse obeys the scaling relation
\begin{equation} \label{mass scaling}
m_{BH} \sim |p-p^*|^\gamma,
\end{equation}
where the scaling exponent, $\gamma$, like the echoing exponent, is universal.  This scaling relation indicates that in type II collapse a black hole can form with arbitrarily small mass.  Gundlach \cite{Gundlach:1996eg} and Hod and Piran \cite{Hod:1996az} showed that the scaling relation (\ref{mass scaling}) is not a strict proportionality, but on top of the linear relationship is a periodic wiggle with period $\Delta/(2\gamma)$.

The scaling relation (\ref{mass scaling}) applies only to supercritical evolutions, i.e.\ evolutions during which a black hole forms.  Scaling relations for subcritical evolutions, i.e.\ evolutions during which a black hole does not form, are known and offer additional means for determining $\gamma$.  Garfinkle and Duncan \cite{Garfinkle:1998va} showed that the maximum value over the total evolution of the central value of the Ricci scalar can obey
\begin{equation} \label{Ricci scaling law}
\mathcal{R}_1 \equiv 
\underset{t}{\text{max}} \, R\indices{^\mu_\mu}(t,0)
\sim |p-p^*|^{-2\gamma},
\end{equation}
where $R_{\mu\nu}$ is the Ricci tensor and $R\indices{^\mu_\mu}$ is the Ricci scalar.  In some systems the above formula is not useful.  For example, in the Einstein-Yang-Mills system with $SU(2)$ the Ricci scalar vanishes.  Garfinkle and Duncan suggested other possibilities \cite{Garfinkle:1998va}, the simplest of which is
\begin{equation} \label{R2 def}
\mathcal{R}_2 \equiv 
\underset{t}{\text{max}} \, |R^{\mu\nu}(t,0)R_{\mu\nu}(t,0)|^{1/2}
\sim |p-p^*|^{-2\gamma}.
\end{equation}
They further argued that the scaling relations (\ref{Ricci scaling law}) and (\ref{R2 def}), like the black hole mass scaling (\ref{mass scaling}), should have a periodic wiggle on top of the linear relationship, again with period $\Delta/(2\gamma)$.

In addition to type II, there is type I and type III critical phenomena.  In type I, originally discovered by Choptuik, Chmaj, and Bizo\'n in their study of gravitational $SU(2)$ \cite{Choptuik:1996yg}, one again considers single-parameter initial data that evolves either to a spacetime containing a black hole or to one that does not.  In this case, however, the black hole must form with finite mass.  Further, the critical solution is a static gravitational solution with a single decay mode.  For example, for the $SU(2)$ system studied in \cite{Choptuik:1996yg}, the critical solution is the $n=1$ Bartnik-McKinnon solution \cite{Bartnik:1988am}.  The closer the evolution is to the critical solution, i.e.\ the closer $p$ is to $p^*$, the longer the evolution spends near the static critical solution before evolving away to one of its two possible end states.

Type III critical phenomena was discovered by Choptuik, Hirschmann, and Marsa \cite{Choptuik:1999gh}, again in a study of gravitational $SU(2)$.  In this case, both end states of the evolution contain a black hole, but the final system is distinctly different depending on whether $p > p^*$ or $p<p^*$.  Type III shares similarities with type I in that the critical solution is a static gravitational solution with a single decay mode (but in this case the static solution contains a black hole) and the closer the evolution is to the critical solution, the longer the evolution spends near the static solution before evolving away to one of its two possible end states.  For $SU(2)$ \cite{Choptuik:1999gh, Rinne:2014kka}, the critical solutions are the colored black hole static solutions \cite{Volkov:1989fi, Bizon:1990sr, Kunzle:1990is}.  For reviews of gravitational critical phenomena see those by Gundlach et al.\ \cite{Gundlach:2002sx, Gundlach:2007gc} and for studies of the critical behavior of gravitational $SU(2)$ see \cite{Choptuik:1996yg, Gundlach:1996je, Choptuik:1999gh, Millward:2002pk, Rinne:2013qc, Rinne:2014kka, Maliborski:2017jyf, kain, Jackson:2018lzc}.

In this paper I study the gravitating 't Hooft-Polyakov magnetic monopole system:\ spherically symmetric $SU(2)$ with a scalar field in the adjoint representation coupled to gravity \cite{tHooft:1974kcl, Polyakov:1974ek, VanNieuwenhuizen:1975tc}.  I previously studied this Einstein-Yang-Mills-Higgs system in \cite{kain} with respect to type III critical behavior.  The well-known solutions for static gravitating monopoles \cite{Lee:1991vy, Ortiz:1991eu, Breitenlohner:1991aa, Breitenlohner:1994di} include both stable and unstable black hole monopole solutions and the Reissner-Nordstr\"om solution.  I showed in \cite{kain} that the unstable static black hole monopole solutions are type III critical solutions with the stable static black hole monopole solutions and the static Reissner-Nordstr\"om solution as the two possible end states.

There exist regular solutions for excited static gravitating monopoles \cite{Breitenlohner:1991aa} which are expected to be unstable.  Further, if the vacuum value of the scalar field is sufficiently large, a branch of unstable fundamental regular static solutions appear \cite{Hollmann:1994fm}.  Both of these are good candidates for a type I critical solution and it would be interesting to study type I critical phenomena in this system.

My focus in this work is on type II critical behavior of gravitating monopoles.  This system offers interesting differences compared to type II collapse found in other systems.  For example, instead of the critical solution sitting between a black hole and complete dispersal of the matter fields, the matter fields on the non-black hole side do not completely disperse, but instead settle down to a stable and static gravitating monopole \cite{Hollmann:1994fm, kain}.

More interesting is that the monopole system appears to contain two type II critical solutions, each with their own set of scaling and echoing exponents.  I note, however, that one solution is more exact than the other.  For the solution I present first, near-critical evolutions exhibit precise self-similarity and, within the scope of initial data that leads to the critical solution, universal scaling and echoing exponents.  The second solution has all the standard signs for a type II critical solution, but the scaling and echoing exponents have a small spread in values over different initial data and the self-similarity of near-critical evolutions is not as precise.  For this second solution, then, it might be that exact self-similarity and universality are lost, a possibility that has also been seen recently in pure $SU(2)$ by Maliborski and Rinne \cite{Maliborski:2017jyf}.

In the next section I present equations, boundary conditions, and aspects of the code I use to study type II collapse.  In Sec.\ \ref{sec:critical 1} I present the first critical solution and in Sec.\ \ref{sec:critical 2} I present the second critical solution.  I conclude in Sec.\ \ref{sec:conclusion}.

%===============================================

\section{Equations, Boundary Conditions, and Numerics}

I gave the full set of equations for the spherically symmetric gravitating monopole in \cite{kain}.  I quickly list the equations here and I refer the reader to \cite{kain} for additional information.  All results will be presented in radial-polar gauge.  This gauge has been used in many studies of type II critical phenomena, including the original study \cite{Choptuik:1992jv} and the first study of pure $SU(2)$ \cite{Choptuik:1996yg}.  In this gauge, the spherically symmetric metric takes a particularly simple form:
\begin{align} \label{spherical metric}
ds^2 
&= - \alpha^2 dt^2 
+ a^2 dr^2
+ r^2 \left(d\theta^2 + \sin^2\theta  d\phi^2 \right)
\end{align}
(here and throughout I set $c=1$), where the metric is parametrized in terms of the lapse $\alpha(t,r)$ and the metric function $a(t,r)$.

The matter sector contains two fields:\ a real scalar field, $\varphi$, which parametrizes the real triplet scalar field gauged under $SU(2)$, and what is effectively a real scalar field, $w$, which parametrizes the gauge field.  That there is only one field parametrizing the gauge field is because the monopole system is within what is called the magnetic ansatz (see, for example, \cite{Choptuik:1999gh,kain} for details).   For simplicity I shall refer to $\varphi$ as the scalar field and $w$ as the gauge field.  From these follow the auxiliary fields:
\begin{align} 
\Phi(t,r) &= \partial_r \varphi&
\Pi(t,r) &=  
\frac{a}{\alpha}\partial_t\varphi
\notag \\
Q(t,r) &= \partial_r w&
P(t,r) &=   \frac{a}{\alpha} 
\partial_t w .
\end{align}

From the Einstein field equations, the metric functions obey the constraint equations
\begin{equation} \label{radial-polar metric equations}
\begin{split}
\frac{\partial_r a}{a} &= 
4\pi G r a^2\rho 
- \frac{a^2 - 1}{2r}
\\
\frac{\partial_r \alpha}{\alpha} &=
4\pi G r  a^2 S\indices{^r_r} +\frac{a^2-1}{2r},
\end{split}
\end{equation}
where $G$ is the gravitational constant and
\begin{align} \label{ADM matter functions}
\rho &=
\frac{\Phi^2 + \Pi^2}{2a^2} 
+ \frac{w^2 \varphi^2}{r^2}
+ V
+ \frac{(1 - w^2)^2 }{2g^2 r^4}
+ \frac{Q^2+ P^2 }{g^2a^2 r^2}
\notag \\
S\indices{^r_r} &=
\frac{\Phi^2 + \Pi^2}{2a^2} 
- \frac{w^2 \varphi^2}{r^2} - V
- \frac{(1 - w^2)^2 }{2g^2  r^4}
+ \frac{Q^2  + P^2  }{g^2a^2 r^2}
\end{align}
follow from the energy-momentum tensor.  $V$ is the scalar potential, which I give below, and $g$ is the gauge coupling constant.  The equations of motion for the matter fields are 
\begin{align} \label{reduced evolution equations}
\partial_t \varphi &= \frac{\alpha}{a} \Pi
\notag \\
\partial_t \Phi &=
\partial_r \left( \frac{\alpha}{a } \Pi\right) 
\notag \\
\partial_t \Pi &= 
\frac{1}{r^2} \partial_r \left( \frac{\alpha  r^2}{a} \Phi \right)
-\alpha a \frac{\partial V}{\partial \varphi}
- \frac{2\alpha a}{r^2} w^2 \varphi 
\notag \\
\partial_t w &= \frac{\alpha}{a} P 
\notag \\
\partial_t Q &=
\partial_r  \left(\frac{\alpha}{a}P \right)
\notag \\
\partial_t P &= \partial_r \left( \frac{\alpha}{a} Q \right)   + \frac{\alpha a}{ r^2}w (1 - w^2 )
- g^2 \alpha a w \varphi^2.
\end{align}

For the matter fields, the inner boundary conditions are
\begin{align} 
\varphi &= O(r)& 
&\Phi = O(1)&
&\Pi = O(r)
\notag \\
w & = 1 + O(r^2)& 
& Q = O(r)&
& P = O(r^2)
\label{inner BC}
\end{align} 
and the outer boundary conditions are $\varphi(t,\infty) = \pm v$, with the rest of the matter fields vanishing at infinity, where $v$ is the vacuum value of the scalar field.  The inner boundary condition for $a$ is $a=1+O(r^2)$.  The inner boundary condition for $\alpha$ is gauge dependent and I shall fix $\alpha = 1$ at the origin, which is a standard gauge choice in studies of type II critical phenomena.

To determine the scaling exponent from the black hole mass scaling law (\ref{mass scaling}), which I'll label as $\gamma_m$, I need to know the black hole mass at the moment of collapse.  The total mass inside a radius $r$ is given by
\begin{equation}
m(t,r) = \frac{r}{2G} \left[1 - \frac{1}{a^2(t,r)} \right],
\end{equation}
which I can use to determine the black hole mass at collapse if I know the horizon radius at collapse.  Since coordinates in radial-polar gauge do not penetrate apparent horizons, I cannot use an apparent horizon finder to find the radius.  As is standard, I take a spike in the metric function $a$ to indicate collapse and its position to be the horizon radius.

The Ricci scalar scaling law (\ref{Ricci scaling law}) is not entirely useful for determining the scaling exponent, which when determined from (\ref{Ricci scaling law}) I'll label as $\gamma_{\mathcal{R}1}$.  I mentioned that the Ricci scalar vanishes in pure $SU(2)$.  Not surprisingly, something similar happens in the gravitating monopole system.  Starting with the Einstein field equations, it is not hard to show that $R\indices{^\mu_\mu}= -8\pi G T\indices{^\mu_\mu}$, where $T_{\mu\nu}$ is the energy-momentum tensor (its components are given in \cite{kain}).  In the gravitating monopole system it can be shown that
\begin{equation}
T\indices{^\mu_\mu} = \frac{\Pi^2 - \Phi^2}{a^2} -  \frac{2w^2 \varphi^2}{r^2}
= -3\Phi^2 + O(r^2),
\end{equation}
where I ignored the scalar potential and where the second equality is only valid near the origin.  We find that the central value of $T\indices{^\mu_\mu}$, and hence also the central value of the Ricci scalar, only probes directly the scalar field and not the gauge field.  Below I shall report the value of $\gamma_{\mathcal{R}1}$, but we should not be surprised if it does not equal $\gamma_m$.

The value of the scaling exponent that follows from the $\mathcal{R}_2$ scaling law (\ref{R2 def}), which I'll label as $\gamma_{\mathcal{R}2}$, is much better adapted for the gravitating monopole system, just as it is for pure $SU(2)$ \cite{Maliborski:2017jyf}.  Starting again from the Einstein field equations, one can show that $R^{\mu\nu}R_{\mu\nu} = (8\pi G)^2 T^{\mu\nu}T_{\mu\nu}$ and further that $T^{\mu\nu}T_{\mu\nu}$ depends explicitly on both $\varphi$ and $w$ near the origin.  The formula for $T^{\mu\nu}T_{\mu\nu}$ is complicated and I do not present it here, but the point is that we should expect $\gamma_{\mathcal{R}2}$ to agree with $\gamma_m$ (at least, for typical type II behavior).

The code I use is the same code used in \cite{kain}, but with three changes.  First, I use radial-polar gauge (instead of radial-maximal gauge) and second, I include Kreiss-Oliger dissipation \cite{AlcubierreBook} to help with stability.  The third, and most important, change has to do with the computational grid.  Finding a type II critical solution requires code that can probe very close to the origin.   The usual best method for doing this is an adaptive mesh \cite{Choptuik:1992jv, Choptuik:1996yg}, but this can be challenging to implement.  A simpler alternative is to use a fixed, but nonuniform computational grid (for examples, see \cite{Akbarian:2015oaa, Maliborski:2017jyf}).  I use the  nonuniform grid used by Akbarian and Choptuik in \cite{Akbarian:2015oaa}:
\begin{equation}
r = e^x - e^{x_\text{min}} + \frac{x_\text{max}}{x_\text{max} - x} - \frac{x_\text{max}}{x_\text{max} - x_\text{min}}, 
\end{equation}
which maps the uniform computational domain $x = (x_\text{min},x_\text{max})$ to the nonuniform radial domain $r = (0,\infty)$.  The results in this paper are for $x_\text{min} = -12$ and $x_\text{max} = 4$, which shrinks the innermost grid point by 2 orders of magnitude compared to the uniform grid, and 2011 grid points. 

To test the universality of the scaling and echoing exponents, I use various families of initial data.  Some of the initial data I've used is
\begin{subequations} \label{varphi IC}
\begin{align}
\varphi(0,r) &= v \tanh \left( r/s \right)
\notag \\
&\qquad+ c \frac{r}{r_0} \left[ e^{-(r-r_0)^2/d^2} + e^{-(r+r_0)^2/d^2}\right]
\label{varphi IC gaussian}\\
\varphi(0,r) &= v \frac{(r/s)^3 - r/s}{(r/s)^3 + c}
\label{varphi IC fraction}\\
\varphi(0,r) &=\frac{v}{2}\Biggl\{ 1 - \left[1 + a \left(1 + \frac{b r}{s}\right) e^{-2(r/s)} \right]
\notag \\
&\qquad\qquad
\times  \tanh \left( \frac{r_0-r}{s}\right)
\Biggr\}
\label{varphi IC CHM}
\end{align}
\end{subequations}
and
\begin{subequations} \label{w IC}
\begin{align}
w(0,r) &= 1 - \tanh^2\left( r/s \right)
+ c \left(\frac{r}{r_0}\right)^2 e^{-(r-r_0)^2/d^2}
\label{w IC gaussian}\\
w(0,r) &= \frac{c - (r/s)^2}{c+(r/s)^4}
\label{w IC fraction}\\
w(0,r) &=\frac{1}{2}\Biggl\{ 1 +  \left[1 + a \left(1 + \frac{b r}{s}\right) e^{-2(r/s)^2} \right]
\notag \\
&\qquad\qquad
\times
 \tanh \left( \frac{r_0-r}{s}\right)
\Biggr\},
\label{w IC CHM}
\end{align}
\end{subequations}
along with $\partial_t \varphi(0,r) = \partial_t w(0,r) = 0$.  In the above equations $s$, $r_0$, $c$, and $d$ are constants and $a$ and $b$ are chosen such that the inner boundary conditions are satisfied and are given by $a= \coth(r_0/s)-1$ and $b = \coth(r_0/s)+1$.  Initial data (\ref{varphi IC gaussian}) and (\ref{w IC gaussian}) take simple functions that satisfy the boundary conditions and add to them Gaussians.  Initial data (\ref{varphi IC CHM}) and (\ref{w IC CHM}) are adaptations to the monopole system of initial data used in \cite{Choptuik:1996yg, Choptuik:1999gh}.

The scalar potential for the monopole system is
\begin{equation}
V =\frac{ \lambda}{4} (\varphi^2 - v^2)^2,
\end{equation}
where $\lambda$ is the scalar field self-coupling and $v$ is the scalar field vacuum value.  The constants $\lambda$, $v$, and $g$ parametrize the gravitating monopole system.  It is possible to absorb $g$ into a redefinition of the fields and parameters so that $\lambda/g^2$ and $v$ determine the model and I see no reason not to expect the scaling and echoing exponents to be functions of them.  To reduce this parameter space I consider only $\lambda = 0$, which is not uncommon, and $\bar{v} \equiv \sqrt{4\pi G} v = 0.2$.  I have looked at type II collapse with other values of $\bar{v}$ and found very small variation in the scaling and echoing exponents, but it would be interesting to look more closely at the dependence.

An important check on the code is whether it can reproduce the accepted values for the scaling and echoing exponents for pure $SU(2)$ \cite{Choptuik:1996yg, Gundlach:1996je}.  Setting $\lambda=v=0$ and using initial data with $\varphi=\partial_t\varphi = 0$ forces fields related to the scalar field ($\varphi$, $\Phi$, $\Pi$) to be permanently zero throughout an evolution, reducing the evolution equations in (\ref{reduced evolution equations}) to those for pure $SU(2)$ \cite{Choptuik:1996yg, Choptuik:1999gh}.  This alone is not sufficient because the outer boundary condition in the monopole system is $w(t,\infty)= 0$, while for pure $SU(2)$ it is $w(t,\infty)= \pm 1$, and so slightly different initial data for $w$ is needed.  Using pure $SU(2)$ initial data
\begin{equation}
w(0,r) = 1 + p e^{-[(r-r_0)/s]^2},
\end{equation}
with $(g/\sqrt{4\pi G})r_0 = 3\sqrt{2}$ and $(g/\sqrt{4\pi G})s=\sqrt{2}/4$, along with $\partial_t w(0,r) = 0$, I find $\gamma_m = 0.1939\pm0.0007$, $\gamma_{\mathcal{R}2} = 0.1959 \pm 0.0003$, $\Delta_{\ln r} = 0.736 \pm 0.001$, and $\Delta_\tau = 0.7354 \pm 0.0002$.  These are consistent with the originally computed values $\gamma = 0.20$ and $\Delta = 0.74$ \cite{Choptuik:1996yg} as well as the more refined values $\gamma = 0.1964\pm0.0007$ and $\Delta =0.73784\pm 0.00002$, obtained by directly perturbing the critical solution \cite{Gundlach:1996je}.

From this point forward, all quantities will be their dimensionless versions, defined by $r \rightarrow (g/\sqrt{4\pi G})r$, $t \rightarrow (g/\sqrt{4\pi G})t$, $\varphi \rightarrow \sqrt{4\pi G} \varphi$, $m_{BH} \rightarrow (g\sqrt{G/4\pi}) m_{BH}$, $\mathcal{R}_1 \rightarrow (\sqrt{4\pi G}/g)^2 \mathcal{R}_1$, and $\mathcal{R}_2 \rightarrow  (\sqrt{4\pi G}/g)^2 \mathcal{R}_2$.  I note that $w$ is already dimensionless.

%=============================================

\section{Critical Solution}
\label{sec:critical 1}

To find a critical solution I start with initial data, such as that in (\ref{varphi IC}) and (\ref{w IC}), with one of the parameters taken to be $p$.  I then tune $p$ toward its critical value, $p^*$, through a bisectional search.  All initial data I've tried such that $p$ is a parameter in the initial data for $\varphi$ (and not $w$), such as $p$ being one of the parameters in (\ref{varphi IC}) (and not in (\ref{w IC})), has lead to (nearly) identical scaling and echoing exponents and hence the same critical solution.  Within a limited sector of initial data, then, the scaling and echoing exponents are universal.

\begin{figure}
\centering
\includegraphics[width=3.15in]{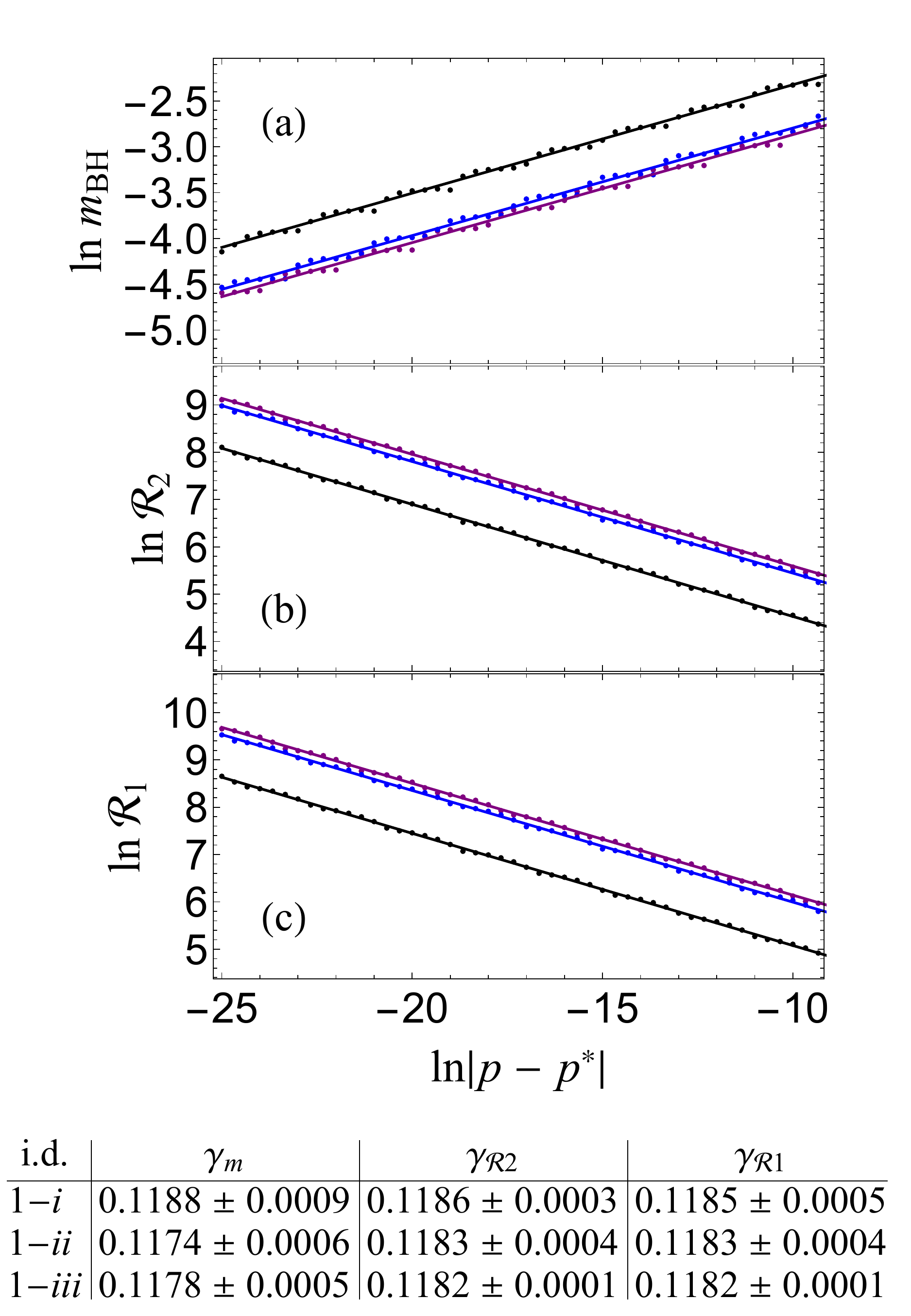}
\caption{Scaling exponent results for three different single-parameter families of initial data:\ Initial data 1-$i$ (black points) is (\ref{varphi IC gaussian}) with $p=c$, $r_0 = 5$, $s=10$, and $d = 0.5$ and (\ref{w IC gaussian}) with $c=0$ and $s=5$; initial data 1-$ii$ (blue points) is (\ref{varphi IC fraction}) with $p=c$ and $s=1$ and (\ref{w IC fraction}) with $c = 2$ and $s=5$; and initial data 1-$iii$ (purple points) is (\ref{varphi IC CHM}) with $p=s$ and $r_0 = 3$ and (\ref{w IC CHM}) with $r_0 = 7$ and $s = 10$.  (a) shows results for the mass scaling relation (\ref{mass scaling}), (b) shows results for the $\mathcal{R}_2$ scaling relation (\ref{R2 def}), and (c) shows results for the  $\mathcal{R}_1$ scaling relation (\ref{Ricci scaling law}).  The table gives the values of the scaling exponents extracted from the best-fit lines.  The scaling exponents appear to agree and be universal.}
\label{fig:varphi scaling}
\end{figure}

Figure \ref{fig:varphi scaling} is a diagram for the scaling exponent.  It displays results using three different single-parameter families of initial data, which I'll label as 1-$i$ (black), 1-$ii$ (blue), and 1-$iii$ (purple), the ``1" indicating that this is initial data for the first critical solution presented.  Plot (a) shows results for the mass scaling relation (\ref{mass scaling}), (b) for the $\mathcal{R}_2$ scaling relation (\ref{R2 def}), and (c) for the $\mathcal{R}_1$ scaling relation (\ref{Ricci scaling law}).  The best-fit lines are determined using a least-squares fit.  Also in Fig.\ \ref{fig:varphi scaling} is a table that gives the values for the scaling exponents $\gamma_m$, $\gamma_{\mathcal{R}2}$, and $\gamma_{\mathcal{R}1}$, as determined from the best-fit lines.  From the table we can see that the different methods for computing the scaling exponent agree with another.  We can also see the universality of the scaling exponent.  Interestingly, $\gamma_{\mathcal{R}1}$ is in agreement with the scaling exponents found using the other methods.  As I explained above, this is not necessarily expected.  It seems to imply that the scalar field dominates over the gauge field in determining the geometry of spacetime at the origin for this critical solution. 

\begin{figure}
\centering
\includegraphics[width=2.5in]{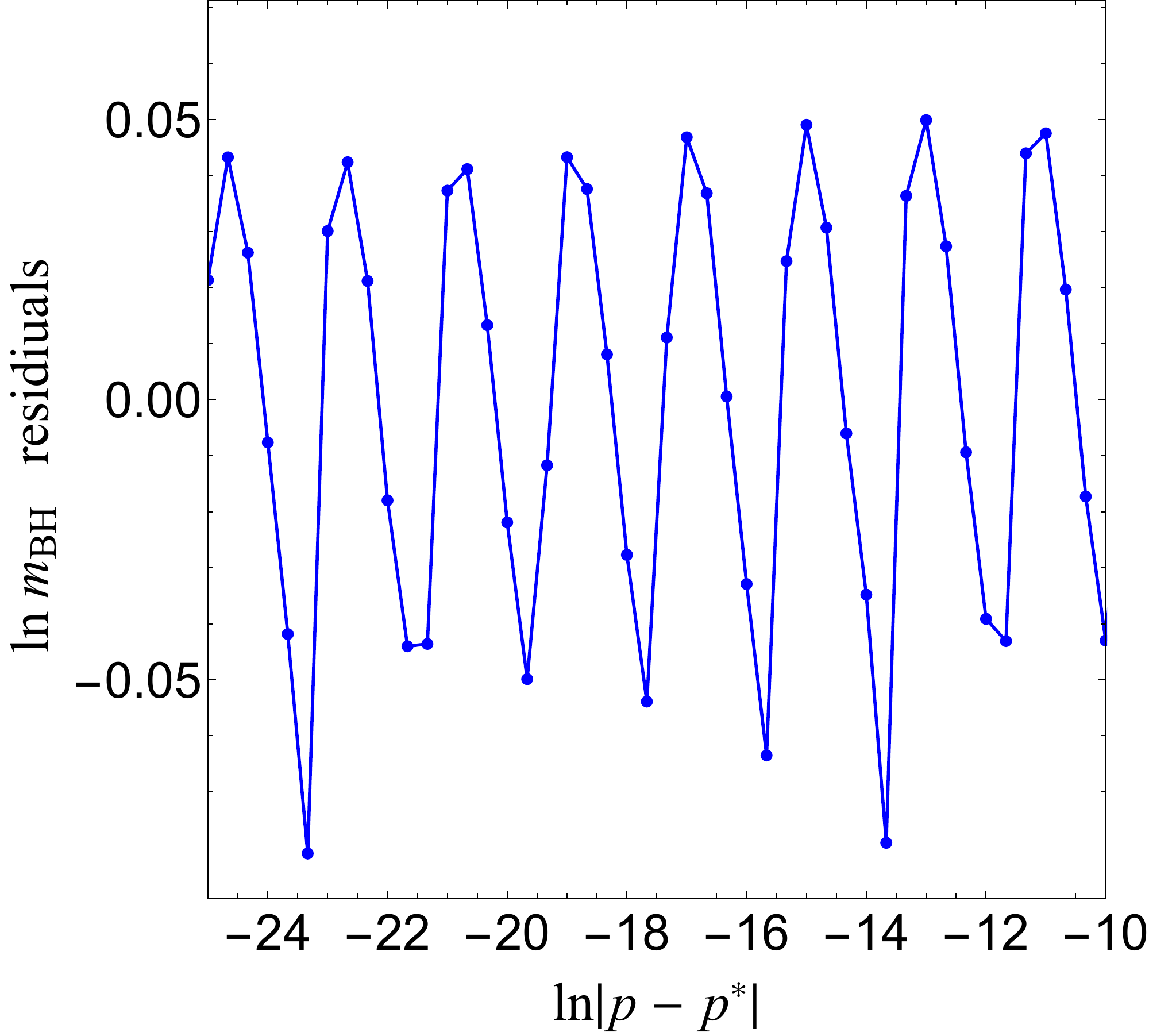}
\caption{Residuals for initial data 1-$ii$ in Fig.\ \ref{fig:varphi scaling}(a).  Each point is found by subtracting from the point in Fig.\ \ref{fig:varphi scaling}(a) the corresponding value of the best-fit line.  The periodicity is clearly seen, with a period right around 2.  This is consistent with $\Delta/(2\gamma) = 1.93$, as computed using the average values of $\gamma$ and $\Delta$ from the tables in Fig.\ \ref{fig:varphi scaling} and Fig.\ \ref{fig:varphi ss} below.  (Note that the lines connecting the points are simple straight lines and are not from any sort of fit.)}
\label{fig:varphi res}
\end{figure}

Even from visual inspection of Fig.\ \ref{fig:varphi scaling}, one can see that all curves have a periodic wiggle around their best-fit line and that the period is roughly equal to 2.  In Fig.\ \ref{fig:varphi res} I show a plot of the residuals for one of the curves in Fig.\ \ref{fig:varphi scaling}(a).  A period of right around 2 is easily seen.  (The Fourier transform of the residuals has a peak at 2, but unfortunately there is not enough data for the Fourier transform to give a more accurate answer.)  In looking at residuals I have found that all scaling data has a period of about 2.  Such a period is consistent with $\Delta/(2\gamma) = 1.93$, where I used the average values of $\gamma$ and $\Delta$ from the tables in Figs.\ \ref{fig:varphi scaling} and \ref{fig:varphi ss}.

\begin{figure}
\centering
\includegraphics[width=3.1in]{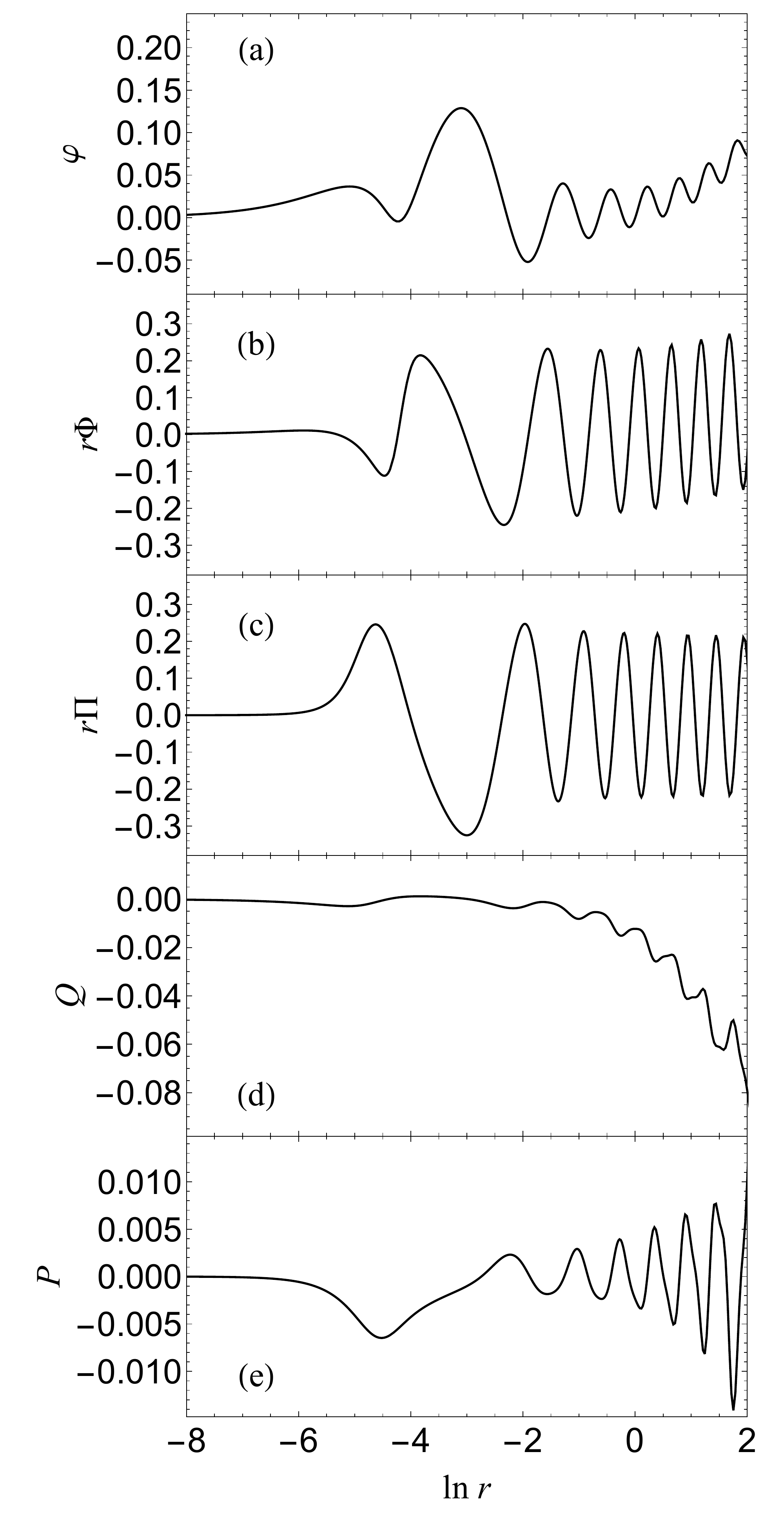}
\caption{Values of five fields for a near-critical evolution at moments in time when the spacetime is on the verge of collapse is shown for initial data 1-$i$.  In (a)--(c), which displays fields associated with the scalar field, we see echoing typical of a type II critical solution.  In (d) and (e), which shows fields associated with the gauge field, we see echoes, but they have a somewhat different appearance.}
\label{fig:varphi echos}
\end{figure}

\begin{figure}
\centering
\includegraphics[width=3.1in]{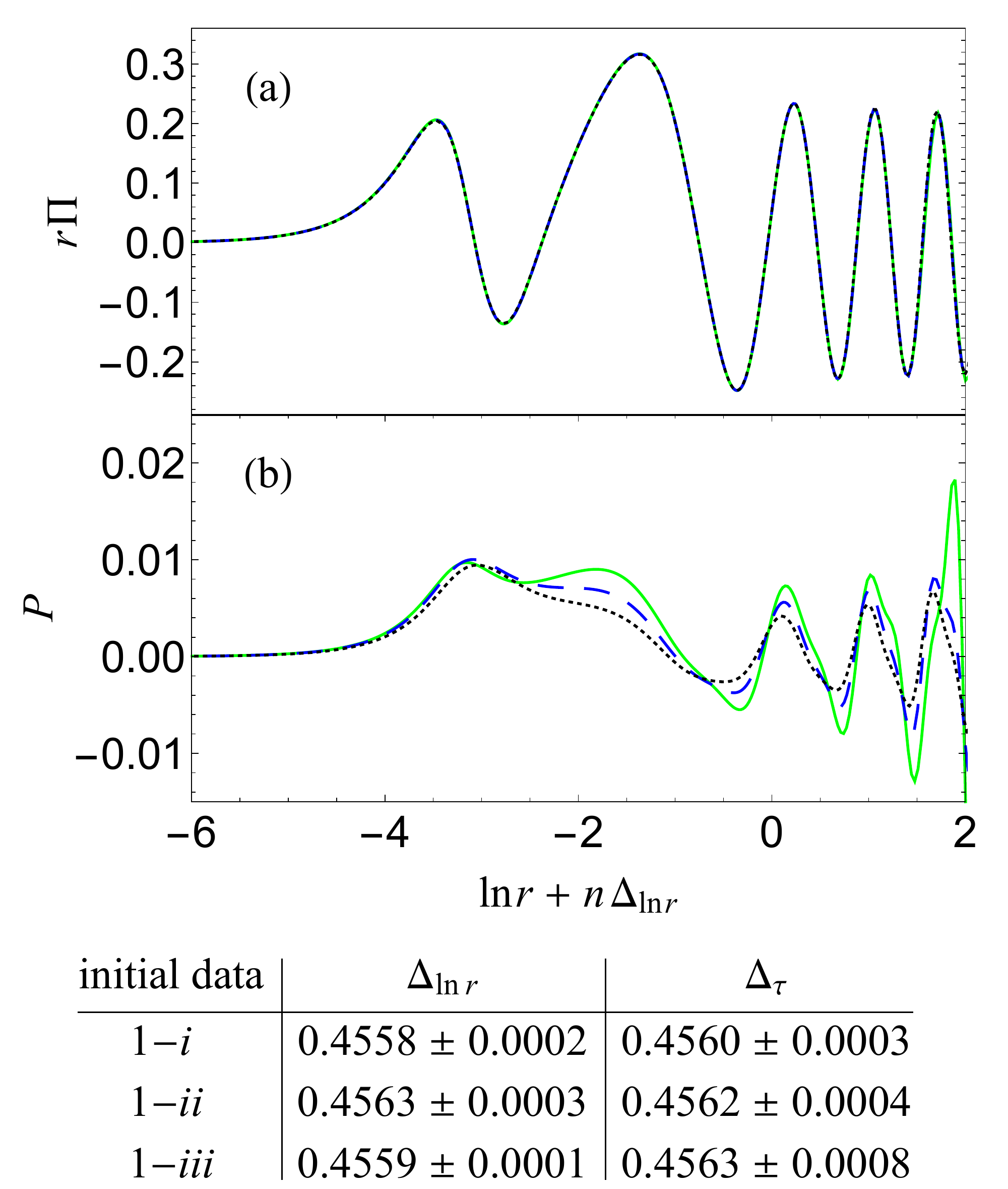}
\caption{(a) and (b) are self-similarity diagrams for the same evolution shown in Fig.\ \ref{fig:varphi echos}, which uses initial data 1-$i$.  $r\Pi$ is a field associated with the scalar field and in (a) we see that it exhibits self-similarity typical of type II behavior, with $n=0$ (solid green), $n=1$ (dashed blue), and $n=2$ (dotted black).  Though not shown, the other fields associated with the scalar field ($\varphi$ and $r\Phi$) also exhibit self-similarity.   $P$ is a field associated with the gauge field and in (b) we see that it does not exhibit self-similarity (nor does the other field associated with the gauge field, $Q$, which is not shown).  The table gives the echoing exponents found for $r\Pi$ for the three families of initial data listed in the caption to Fig.\ \ref{fig:varphi scaling}.  The echoing exponents appear to agree and be universal.}
\label{fig:varphi ss}
\end{figure}

Figure \ref{fig:varphi echos} displays a near-critical evolution, with $\ln|p-p^*| \approx -32$ (or $|p-p^*| \approx 10^{-14}$), using initial data 1-$i$, and is plotted at moments in time when the spacetime is on the verge of collapse.  The top three figures, (a)--(c), plot fields associated with the scalar field and the echoing is readily seen to be typical of a type II critical solution.  The bottom two figures, (d) and (e), plot fields associated with the gauge field, but the echoing has a somewhat different appearance.

In Fig.\ \ref{fig:varphi ss} I show diagrams whose purpose is to exhibit the discrete self-similarity of the solutions, if it exists.  I shall refer to these diagrams, and the analogous ones below in Fig.\ \ref{fig:w ss}, as self-similarity diagrams.  In such diagrams I plot a near-critical scale invariant function, $Z$, at some central proper time, $T$, which, in terms of coordinate time $t$, is given by
\begin{equation}
T(t) = \int_0^t \alpha(t',0) dt'.
\end{equation}
I then search for values of $\Delta_{\ln r}$ and $\Delta_\tau$ such that $Z(\tau+n\Delta_\tau, \ln r + n \Delta_{\ln r})$ and $Z(\tau,\ln r)$ overlap, where $n$ is a positive integer, $\tau = \ln(T^*-T)$, and $T^*$ is a constant called the accumulation time.  The expectation in type II collapse is that $\Delta_{\ln r} = \Delta_\tau$.

Figure \ref{fig:varphi ss}(a) displays the discrete self-similarity of the field $r\Pi$.  Though not shown, the other fields associated with the scalar field ($\varphi$ and $r\Phi$) also exhibit self-similarity analogously to Fig.\ \ref{fig:varphi ss}(a).  The table in Fig.\ \ref{fig:varphi ss} gives the echoing exponents found for $r\Pi$ for the three families of initial data listed in the caption to Fig.\ \ref{fig:varphi scaling}.  The table suggests the echoing exponents are universal and that $\Delta_{\ln r}$ and $\Delta_\tau$ agree.

Figure \ref{fig:varphi ss}(b) shows a self-similarity diagram for $P$, a field associated with the gauge field.  The curves in Fig.\ \ref{fig:varphi ss}(b) are for the same times and the same $\Delta_{\ln r}$ used in Fig.\ \ref{fig:varphi ss}(a).  It is clear that the field $P$ is not exhibiting self-similarity.  Though not shown, neither does $Q$.  In general, for the critical solution of this section, there does not exist values for $\Delta_{\ln r}$ and $\Delta_\tau$ such that the fields associated with the gauge field ($Q$ and $P$) exhibit self-similarity.

%=============================================

\section{Second Critical Solution}
\label{sec:critical 2}

Almost all initial data I've tried such that $p$ is a parameter in the initial data for $w$ (and not $\varphi$) has lead to, by all appearances, a second critical solution, with scaling and echoing exponents different from those in the previous section.  However, the scaling and echoing exponents are not as universal and the self-similarity is not as precise as in the previous section.

\begin{figure}
\centering
\includegraphics[width=3.15in]{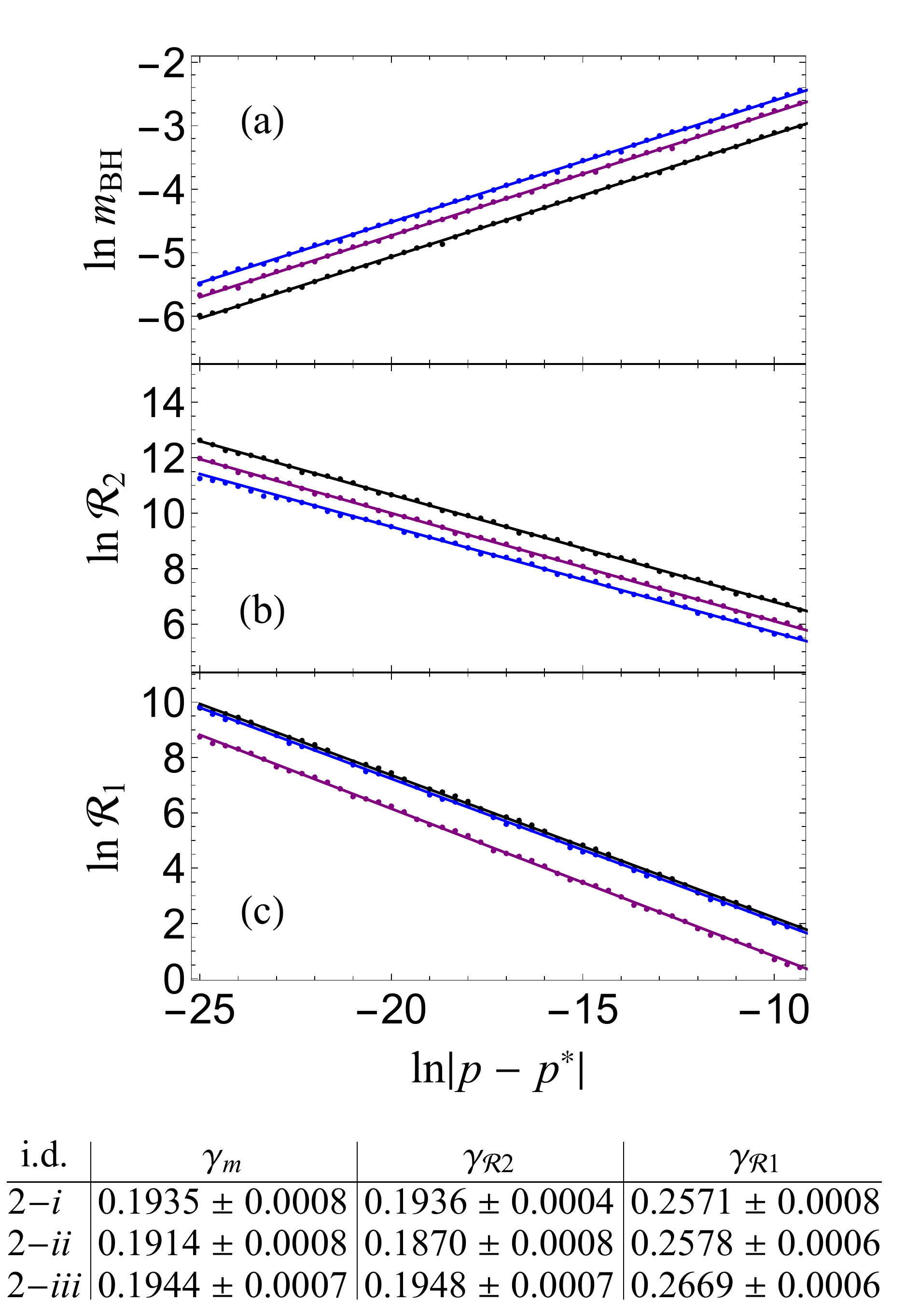}
\caption{Scaling exponent results for three different single-parameter families of initial data:\ Initial data 2-$i$ (black points) is (\ref{varphi IC gaussian}) with $c=0$ and $s=5$ and (\ref{w IC gaussian}) with $p = c$, $s=10$, and $r_0 = 5$, and $d=0.5$; initial data 2-$ii$ (blue points) is (\ref{varphi IC fraction}) with $c=3$ and $s=1$ and (\ref{w IC fraction}) with $p=c$ and $s=4$; and initial data 2-$iii$ (purple points) is (\ref{varphi IC CHM}) with $r_0 = 7$ and $s=10$ and (\ref{w IC CHM}) with $s=p$ and $r_0 = 3$.  (a) shows results for the mass scaling relation (\ref{mass scaling}), (b) shows results for the $\mathcal{R}_2$ scaling relation (\ref{R2 def}), and (c) shows results for the $\mathcal{R}_1$ scaling relation (\ref{Ricci scaling law}).  The table gives the values of the scaling exponents extracted from the best-fit lines.  Comparing this to Fig.\ \ref{fig:varphi scaling}, we see that the scaling exponent for the critical solution of this section is not as universal as the scaling exponent for the critical solution of the previous section.  I note that $\gamma_{R1}$ does not equal $\gamma$ as found from the other methods, which, as explained in the main text, is not unexpected.}
\label{fig:w scaling}
\end{figure}

Figure \ref{fig:w scaling} is a diagram for the scaling exponent.  It displays results using three different single-parameter families of initial data (which are different than those used in the previous section), which I'll label as 2-$i$ (black), 2-$ii$ (blue), and 2-$iii$ (purple), the ``2" indicating that this is initial data for the second critical solution presented.  Plot (a) shows results for the mass scaling relation (\ref{mass scaling}), (b) for the $\mathcal{R}_2$ scaling relation (\ref{R2 def}), and (c) for the $\mathcal{R}_1$ scaling relation (\ref{Ricci scaling law}).  The best-fit lines are determined using a least-squares fit.  Also in Fig.\ \ref{fig:w scaling} is a table that gives the values for the scaling exponents $\gamma_m$, $\gamma_{\mathcal{R}2}$, and $\gamma_{\mathcal{R}1}$, as determined from the best fit lines.  From the table we can see that the scaling exponent is not as universal for the critical solution of this section as the scaling exponent is for the critical solution of the previous section.

For the critical solution of this section I find that $\gamma_{\mathcal{R}1}$ does not agree with the scaling exponent found using the other methods.  As I explained above, this is not unexpected.  This implies that the gauge field is playing a much more important role in the critical solution of this section, something that is also not unexpected.  Interestingly, however, the value of $\gamma_{\mathcal{R}1}$, though not in agreement with $\gamma_m$ and $\gamma_{\mathcal{R}2}$, is as universal in its value as they are in their values. 

\begin{figure}
\centering
\includegraphics[width=2.5in]{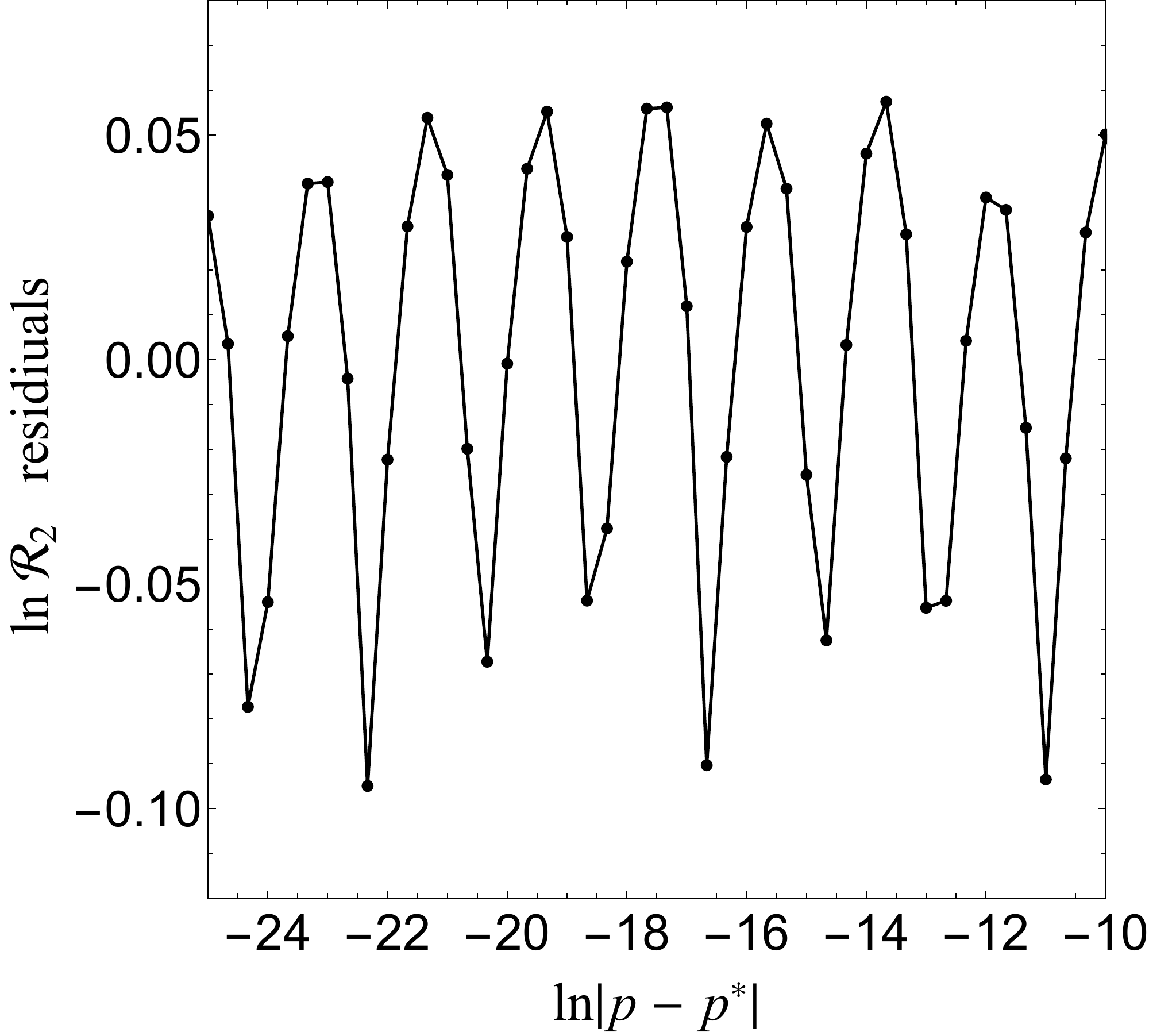}
\caption{Residuals for initial data 2-$i$ in Fig.\ \ref{fig:w scaling}(b).  Each point is found by subtracting from the point in Fig.\ \ref{fig:w scaling}(b) the corresponding value of the best-fit line.  The periodicity is clearly seen, with a period right around 2.   This is consistent with $\Delta/(2\gamma) = 1.89$, as computed using the average values of $\gamma$ and $\Delta$ from the tables in Fig.\ \ref{fig:w scaling} and Fig.\ \ref{fig:w ss} below.  (Note that the lines connecting the points are simple straight lines and are not from any sort of fit.)}
\label{fig:w res}
\end{figure}

A periodic wiggle, though small, can be seen in Fig.\ \ref{fig:w scaling}.  In Fig.\ \ref{fig:w res} I show a plot of the residuals for one of the curves in Fig.\ \ref{fig:varphi scaling}(b) and a period of right around 2 is easily seen.  (The Fourier transform of the residuals has a peak at 2, but unfortunately there is not enough data for the Fourier transform to give a more accurate answer.)  In looking at residuals I have found that all scaling data has a period of about 2.  Such a period is consistent with $\Delta/(2\gamma) = 1.89$, where I used the average values of $\gamma$ and $\Delta$ from the tables in Figs.\ \ref{fig:w scaling} and \ref{fig:w ss}.

\begin{figure}
\centering
\includegraphics[width=3.1in]{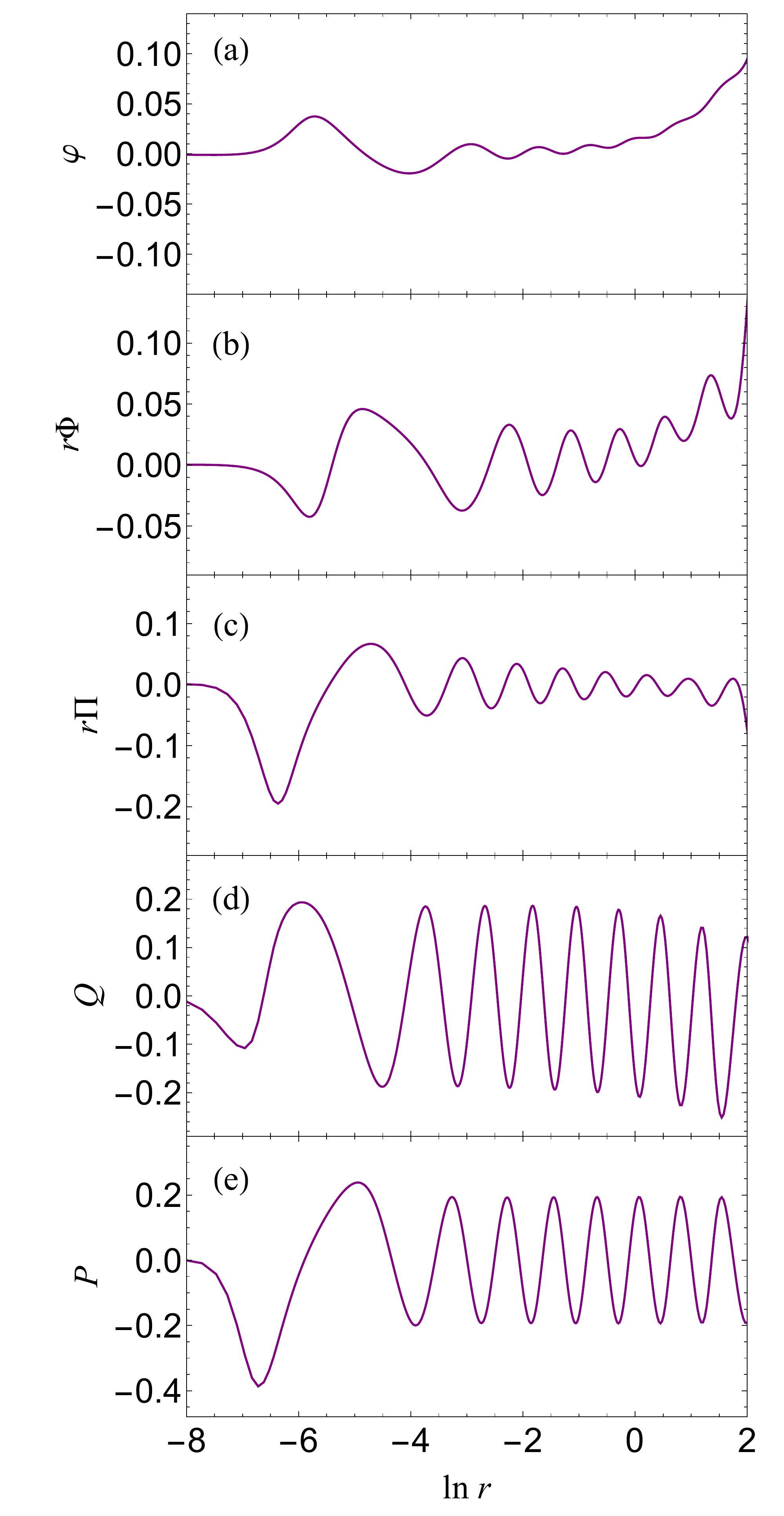}
\caption{Values of five fields for a near-critical evolution at moments in time when the spacetime is on the verge of collapse is shown for initial data 2-$iii$.  Comparing with Fig.\ \ref{fig:varphi echos}, we see that for the critical solution of this section, it is instead the fields associated with the gauge field in (d) and (e) that exhibit echoing typical of a type II critical solution and it is the fields associated with the scalar field in (a)--(c) that have a somewhat different appearance.}
\label{fig:w echos}
\end{figure}

\begin{figure}
\centering
\includegraphics[width=3.1in]{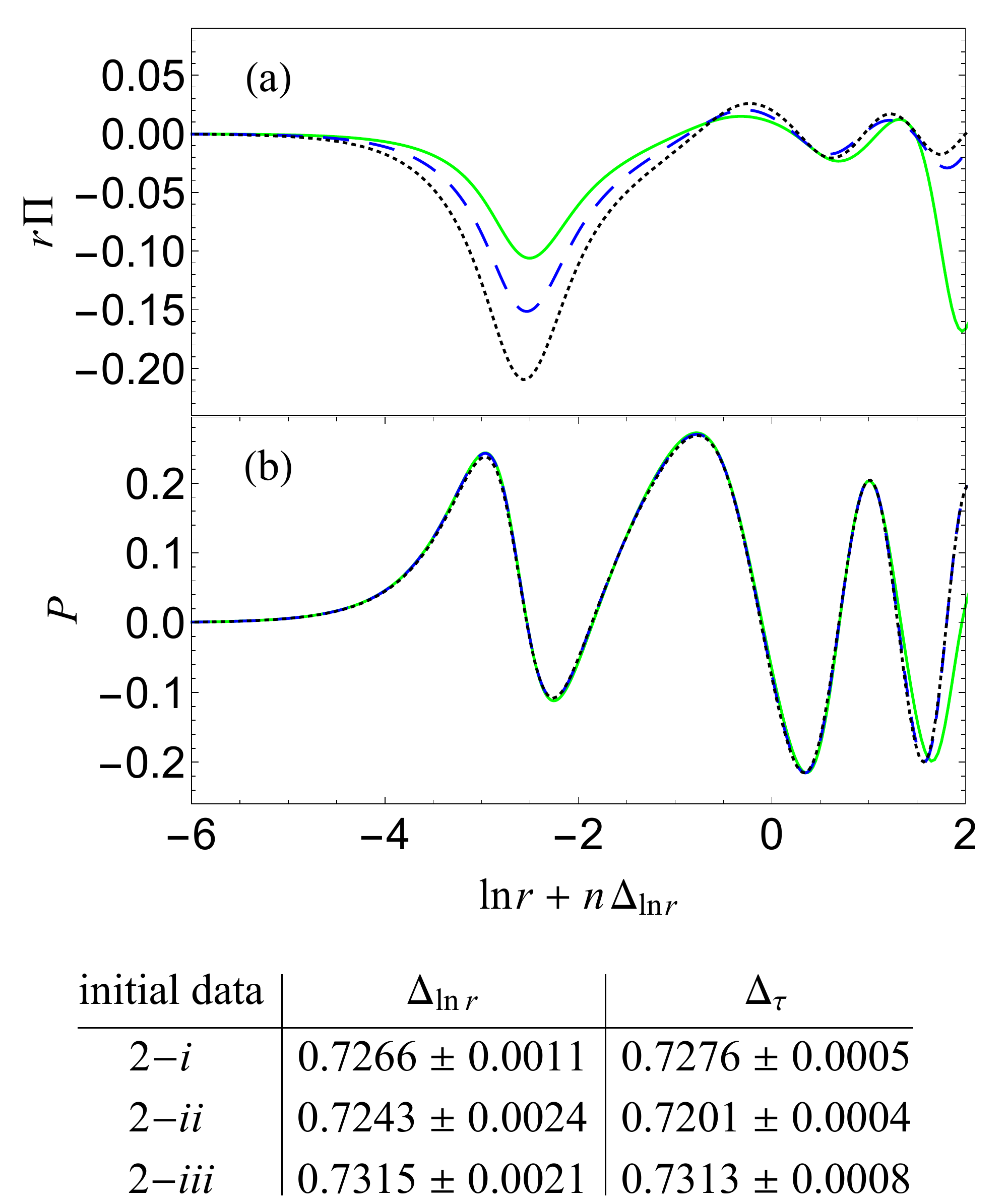}
\caption{(a) and (b) are self-similarity diagrams for the same evolution shown in Fig.\ \ref{fig:w echos}, which uses initial data 2-$iii$.  $P$ is a field associated with the gauge field and in (b) we see that it exhibits self-similarity typical of type II behavior with $n=0$ (solid green), $n=1$ (dashed blue), and $n=2$ (dotted black).  Though not shown, the other field associated with the gauge field ($Q$) also exhibits self-similarity.  $r\Pi$ is a field associated with the scalar field and in (a) we see that it does not exhibit self-similarity (nor do the other fields associated with the scalar field, $\varphi$ and $r\Phi$, which are not shown).  The table gives the echoing exponents for $P$ for the three families of initial data listed in the caption of Fig.\ \ref{fig:w scaling}.  From the table we see that the echoing exponent is not as universal for the critical solution of this section as it is for the critical solution of the previous section.}
\label{fig:w ss}
\end{figure}

Figure \ref{fig:w echos} displays a near-critical evolution, with $\ln|p-p^*| \approx -32$ (or $|p-p^*| \approx 10^{-14}$), using initial data 2-$iii$, and is plotted at moments in time when the spacetime is on the verge of collapse.  The top three figures, (a)--(c), plot fields associated with the scalar field and the bottom two figures, (d) and (e), plot fields associated with the gauge field.  Comparing with Fig.\ \ref{fig:varphi echos}, we see that for the critical solution of this section, it is instead the fields associated with the gauge field that exhibit echoing typical of a type II critical solution and it is the fields associated with the scalar field that have the somewhat different appearance. 

Figure \ref{fig:w ss} displays discrete self-similarity diagrams for $r\Pi$ and $P$, and may be compared with Fig.\ \ref{fig:varphi ss}.  We find now that it is the field associated with the gauge field, $P$ in Fig.\ \ref{fig:w ss}(b), which exhibits self-similarity (though not shown, $Q$ exhibits it as well), and it is the field associated with the scalar field, $r\Pi$ in Fig.\ \ref{fig:w ss}(a), which does not (though not shown, neither does $\varphi$ nor $r\Phi$).  There does not exist values for $\Delta_{\ln r}$ and $\Delta_\tau$ such that the fields associated with the scalar field ($\varphi$, $r\Phi$, and $r\Pi$) exhibit self-similarity.  The table in Fig.\ \ref{fig:w ss} gives echoing exponents for the three families of initial data listed in the caption of Fig.\ \ref{fig:w scaling}.  Just as with the scaling exponent, the table shows us that the echoing exponent, $\Delta$, is not as universal for the critical solution of this section as it is for the critical solution of the previous section.  Further, close inspection of Fig.\ \ref{fig:w ss}(b) and analogous diagrams made with different initial data shows that self-similarity is not as exact for the critical solution of this section as it is for the critical solution of the previous section.

It would be elegant if all initial data such that $p$ is a parameter in the initial data for $w$ led to the critical solution of this section.  Though this is the case for nearly all initial data I've tried, I have found exceptions.  For example, initial data (\ref{varphi IC fraction}) with $c=2$ and $s = 5$ and (\ref{w IC fraction}) with $c=p$ and $s=1$ leads to the critical solution of the previous section.

It is unclear why self-similarity is less exact and the scaling and echoing exponents are less universal for the critical solution of this section compared to the critical solution of the previous section.  It is impossible to completely rule out this being a numerical artifact, but I have found no evidence for this.  It may very well be, that for the critical solution of this section, exact self-similarity and universality are lost.  Intriguingly, something similar was seen recently by Maliborski and Rinne in their study of type II critical behavior in pure $SU(2)$ \cite{Maliborski:2017jyf}.  It may be useful to touch on the similarities of their system and the system studied here.  Most numerical studies of gravitational $SU(2)$, including the original studies \cite{Choptuik:1996yg, Choptuik:1999gh} , work within the magnetic ansatz (the present work is also within the magnetic ansatz), which reduces the four fields parametrizing the spherically symmetric $SU(2)$ gauge field down to a single field.  Maliborski and Rinne \cite{Maliborski:2017jyf} are the first to study critical behavior in $SU(2)$ without making the magnetic ansatz.  The particular $SU(2)$ gauge they work in reduces the four gauge fields down to effectively two fields (there is a third field, but it obeys a constraint equation instead of an equation of motion).  Beyond the fact that both the system studied in \cite{Maliborski:2017jyf} and the present system are part of $SU(2)$, an obvious similarity is that both systems have multiple matter fields.  It would be interesting to know what role, if any, this plays in the possible loss of universality and self-similarity.

%=============================================

\section{Conclusion}
\label{sec:conclusion}

In this work I studied type II critical behavior in the gravitating magnetic monopole system.  This system is characterized by two matter fields:\ A real scalar field, which parametrizes the scalar field gauged under $SU(2)$, and what is effectively a real scalar field, which parametrizes the gauge field.  This system offers some differences compared to other systems.  For example, on the non-black hole side of the critical solution, the matter fields do not completely disperse, but instead settle down to a stable and static configuration.  More interesting, however, is that the gravitating monopole system appears to have two critical solutions.  

All initial data I tried in which the scalar field is tuned toward a critical value led to the critical solution I presented first in Sec.\ \ref{sec:critical 1}.  This critical solution exhibits precise self-similarity and universal scaling and echoing exponents.  In Sec.\ \ref{sec:critical 2} I presented a second critical solution, which most of the initial data I tried in which the gauge field is tuned toward a critical value led to, but I did find exceptions.  Though this second critical solution has different scaling and echoing exponents than the first critical solution, the self-similarity is less exact and the scaling and echoing exponents are less universal.  Indeed, exact self-similarity and universality of the scaling and echoing exponents may be lost, a possibility that was recently seen elsewhere \cite{Maliborski:2017jyf}.

It is interesting that, in the first critical solution of Sec.\ \ref{sec:critical 1}, which is obtained by tuning the scalar field toward a critical value, the fields associated with the scalar field exhibit self-similarity, while the fields associated with the gauge field do not.  And on the other hand, in the second critical solution of Sec.\ \ref{sec:critical 2}, which is usually obtained by tuning the gauge field toward a critical value, it flips, with the fields associated with the gauge field exhibiting self-similarity, while the fields associated with the scalar field do not.

%====================================================================

%\bibliography{typeII}


\end{document}